\documentclass[10pt,twocolumn,letterpaper]{article}

\usepackage[pagenumbers]{cvpr} % To force page numbers, e.g. for an arXiv version

\usepackage{graphicx}
\usepackage{amsmath}
\usepackage{amssymb}
\usepackage{booktabs}
\usepackage{algorithm}
\usepackage{algorithmic}
\usepackage{amsmath}
\usepackage[accsupp]{axessibility}

\usepackage[pagebackref,breaklinks,colorlinks]{hyperref}

\usepackage[capitalize]{cleveref}
\crefname{section}{Sec.}{Secs.}
\Crefname{section}{Section}{Sections}
\Crefname{table}{Table}{Tables}
\crefname{table}{Tab.}{Tabs.}

\def\cvprPaperID{3994} % *** Enter the CVPR Paper ID here
\def\confName{CVPR}
\def\confYear{2022}

\begin{document}

\title{Few-shot Backdoor Defense Using Shapley Estimation}

\author{Jiyang Guan$^{1,2}$\thanks{This work was done when he was research intern at JDEA}, ~Zhuozhuo Tu$^{3*}$, ~Ran He$^{1,2}$\thanks{Corresponding author}, ~Dacheng Tao$^{4,3}$\\
$^{1}$NLPR $\& $CRIPAC, Institute of Automation, Chinese Academy of Sciences, China \\
$^{2}$School of Artificial Intelligence, University of Chinese Academy of Sciences, China\\
$^{3}$The University of Sydney, Australia\\
$^{4}$JD Explore Academy, China\\
{\tt\small guanjiyang2020@ia.ac.cn,
\tt\small zhtu3055@uni.sydney.edu.au,}\\
{\tt\small rhe@nlpr.ia.ac.cn,
\tt\small dacheng.tao@gmail.com
}
}

\maketitle

%%%%%%%%% ABSTRACT
\begin{abstract}
   Deep neural networks have achieved impressive performance in a variety of tasks over the last decade, such as autonomous driving, face recognition, and medical diagnosis. However, prior works show that deep neural networks are easily manipulated into specific, attacker-decided behaviors in the inference stage by backdoor attacks which inject malicious small hidden triggers into model training, raising serious security threats. To determine the triggered neurons and protect against backdoor attacks, we exploit Shapley value and develop a new approach called Shapley Pruning (ShapPruning) that successfully mitigates backdoor attacks from models in a data-insufficient situation (1 image per class or even free of data). Considering the interaction between neurons, ShapPruning identifies the few infected neurons (under $1\%$ of all neurons) and manages to protect the model’s structure and accuracy after pruning as many infected neurons as possible. To accelerate ShapPruning, we further propose discarding threshold and $\epsilon$-greedy strategy to accelerate Shapley estimation, making it possible to repair poisoned models with only several minutes. Experiments demonstrate the effectiveness and robustness of our method against various attacks and tasks compared to existing methods.
\end{abstract}

%%%%%%%%% BODY TEXT
\section{Introduction}
\label{sec:intro}
Over the past years, Deep Neural Networks (DNNs) play a great role in machine learning and are applied in many critical domains such as face recognition \cite{taigman2014deepface}, image generation\cite{fu2021high,fu2021dvg}, autonomous driving \cite{el2021leveraging}, and medical diagnosis \cite{kermany2018identifying,yu2021recognizing}. However, because of a lack of transparency and interpretability \cite{mascharka2018transparency,he2009robust,yu2020graph}, DNNs are easy to be manipulated by an adversary into attacker-decided behaviors and make serious mistakes in security-related areas, causing serious threats and concerns. For example, it has been observed that adding deliberate and small distortion to the images in inference stage(\ie, adversarial examples) can cause misclassification in neural network classifiers\cite{goodfellow2014explaining}. 

Backdoor attacks, on the other hand, are a different type of attack, making use of opacity and overfitting of DNNs to create a maliciously trained network which achieves state-of-the-art performance on normal samples but behaves badly on specific attacker-chosen inputs. Gu \etal \cite{gu2017badnets} demonstrates that, compared with adversarial examples, backdoor attacks can cause wrong predictions in models with much smaller distortion. Meanwhile, for black-box models like DNNs, it is difficult to identify the backdoor, and we can only use the test dataset to judge whether they are poisoned. Thus, the backdoor attack is more imperceptible and dangerous \cite{feng2021fiba,azizi2021t}. Furthermore, as training on cloud or directly using the third-party trained models becomes more common today \cite{zhang2020empowering}, backdoor attacks have more access to the models' training procedure. Thus, it is much easier for them to inject triggers into models in recent years.

\begin{figure*}
\centering
\includegraphics[width=0.85\textwidth]{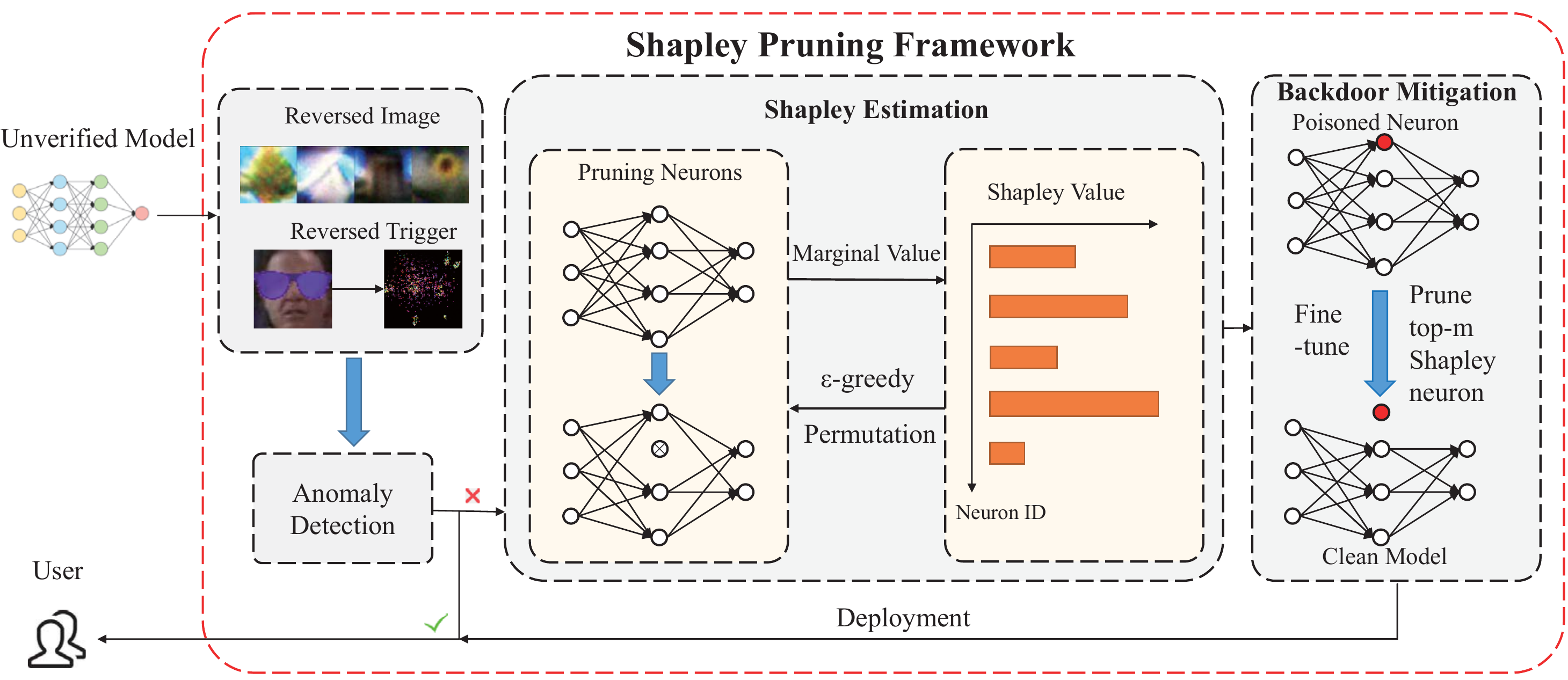}
\centering
\caption{Shapley Pruning Framework. Our framework consists of four components, trigger and data reverse, detection, Shapley estimation, backdoor mitigation, and can effectively remove backdoor in models.}
\label{shapley_pruning}
\end{figure*}

The vulnerabilities to backdoor attacks raise concerns about the security of DNNs \cite{li2020backdoor}, and many defense methods have been proposed, trying to mitigate backdoor from the models \eg Fine Pruning \cite{liu2018fine}, Neural Cleanse \cite{wang2019neural}, GangSweep \cite{zhu2020gangsweep} \etc. However, these methods need a relatively large amount of clean data (\eg $10\%$ of training data required in Neural Cleanse), and can't locate poisoned neurons accurately (\eg pruning 70$\%$ of all neurons in Fine Pruning). To determine the poisoned neurons and mitigate backdoor, we introduce Shapley value and propose a ShapPruning framework to guide detecting the attacked neurons, which successfully mitigates backdoor in the given models. Shapley value is a concept from game theory and is used to allocate worth to cooperative players \cite{shapely1953value,castro2009polynomial,ghorbani2020neuron}. We use Shapley value to attribute the overall backdoor behavior to each neuron and find neurons with the largest Shapley value which are the most responsible for backdoor behavior in models. Compared to prior work, our ShapPruning method can handle the data-insufficient situation and needs only a tiny amount of data (\eg one image per class or even free of clean data) and prunes a very small number of neurons (about $1\%$ of all the neurons), to maintain a good classification accuracy (under $1\%$ accuracy decline in most cases) and clean backdoor clearly.

Our contributions are summarized as follows:
\begin{itemize}
\item We introduce Shapley value into the backdoor area and propose a backdoor mitigation method called Shapley Pruning which can locate and prune poisoned neurons accurately with the reversed trigger.
\item We also propose discarding threshold and $\epsilon$-greedy to accelerate Shapley value's estimation, which yields a more accurate estimation with much less time.
\item Our method considers the relationship between neurons and locates the attacked neurons accurately with few images. As a result, it can prune only $1\%$ of all neurons to recover the model with a small accuracy decrease (accuracy declines $0.1\%$ in the GTSRB dataset and the attack success rate drops to $0.4\%$). Moreover, our method is robust in different situations.

\item We utilize information in model's batch normalization layer and propose a data-free backdoor cleanse method with mixture-mode ShapPruning.

\end{itemize}

\section{Related Work}
Many defense methods have been proposed to deal with the security threats of backdoor attacks. From the perspective of the defender, there are two main settings to mitigate backdoor, \ie, model available defense and data available defense. Data available defenses usually use anomaly detection to detect and eliminate abnormal images in the poisoned training dataset \cite{tran2018spectral,du2019robust}, or weaken the influence of backdoor dataset during model training \cite{rosenfeld2020certified,steinhardt2017certified,li2021anti}. However, in many cases, datasets are unavailable due to privacy concerns and what we can have access to is only trained models which might be injected malicious backdoor attacks. Thus, model available defense attract more attention. Our work considers this setting and focuses on clean data insufficient situations to recover poisoned models. 
    
There is a broad body of literature trying to solve this problem. Fine Pruning \cite{liu2018fine} uses the activation of each neuron on clean data to determine which neurons to prune. But, because deep neural networks are complicated, using activation to guide neuron pruning ignores the correlation between neurons and can't locate the poisoned neurons accurately. Neural Cleanse \cite{wang2019neural} tries to reverse triggers and uses an unlearning way to patch the model. To improve Neural Cleanse, GangSweep \cite{zhu2020gangsweep}, Tabor \cite{guo2020towards}, and DeepInspect \cite{chen2019deepinspect} were proposed to use GANs \cite{goodfellow2014generative} and interpretable AI to generate better-reversed triggers. However, these methods can’t accurately locate the neurons under attack, and their performance, to some extent, depends on fine-tuning. As a result, they usually need a relatively large amount of clean data and prune a large number of neurons. When clean data is insufficient, the performance of these methods declines. Besides, DeepInspect \cite{chen2019deepinspect} is fragile, limited, and the data reverse used by this method is based on a single-layer network and a small face dataset situation \cite{fredrikson2015model}. Unlike the previous method, our method can mitigate backdoor from the poisoned models with only few images (even without clean data) and prune only a few neurons.

\section{Method}
We present our Shapley Pruning framework in this section. Firstly, we introduce Shapley value to DNNs and give its definition. Then, we give an algorithm for estimating Shapley value where we propose $\epsilon$-greedy and discarding threshold to accelerate its estimation. Since Shapley value is evaluated on backdoor dataset, we involve trigger reverse synthesis to generate that dataset. Finally, we involve image recovery and propose a data-free backdoor mitigation method. An overview of our framework is given in \cref{shapley_pruning}.

\subsection{Shapley Value}
In DNNs, since there are a large number of neurons and complicated interactions between them, it is difficult to quantify each neurons' contribution to the overall output. To tackle this problem, we introduce Shapley value which, as one of the most important concepts in cooperative game theory, can allocate values to each player using the average of marginal values \cite{castro2009polynomial}, and be used to determine the contribution of each neuron to the overall output \cite{ghorbani2020neuron}. We can treat a network as an $n$-player game with each neuron as a player. Let $N$ be the set of all $n$ neurons in the neural networks, denoted by $N= \lbrace 1,\ldots,n \rbrace$ and $m$ be a metric function evaluating performance of players. In neural networks, $m$ can be a score function such as accuracy or loss. The marginal contribution of neuron $i$ can be defined as:
\begin{equation}
margin(i)=m(C \cup\{i\})-m(C)  
\end{equation}
where $C$ is a subset of players not containing $i$, i.e., expressed as  $C\subset N\backslash i$. With the marginal contributions, Shapley value $\phi$ for neuron $i$ can be defined using the average of them as follows \cite{shapely1953value}:
\begin{equation}
\phi_{i}(m) = \frac{1}{n} \sum\limits_{C\subset N\backslash i}P_{C}\cdot(m(C\cup i)-m(C))
\label{eq:Shapley}
\end{equation}
where $P_{C}=\frac{(n-c-1)! c!}{(n-1)!}$ represents the relative importance of subset $C$ , and $c$ is the cardinality of $C$. In the next subsection, we will provide an algorithm for computing Shapley value for each neuron.

\subsection{Estimation for Shapley Value}
From \cref{eq:Shapley}, Shapley value can be expressed as the average of marginal contributions of the neuron in all possible orders. We define $O$ as a permutation of neurons and  ${Af}^{i}(O) $ means a subset of neurons after neuron $i$ in the order $O$. $\pi(N)$ is all possible orders of neurons. Then, Shapley value of neuron $i$ can be rewritten as follows \cite{castro2009polynomial}:
\begin{equation}
\begin{aligned}
 \phi_{i}(m)=\sum_{O\in\pi(N)} \frac{1}{n!}(m({Af}^{i}(O) \cup i)&-m({Af}^{i}(O)))\\  i=1, \ldots, n\quad 
 \label{eq:estimation}
\end{aligned}
\end{equation}
\cref{eq:estimation} shows that computing $\phi_{i}$ is equivalent to calculating the expectation of a random variable. Despite that estimating Shapley value exactly is time-consuming as it involves $n!$ permutations of all neurons in deep neural networks, we can approximate it by applying the Monte-Carlo estimation \cite{ferrenberg1989optimized} which first samples permutations of neurons and then calculates the average of marginal contributions with those sampled permutations. Further, we propose discarding threshold and $\epsilon$-greedy acceleration to estimate Shapley value more fast and precisely.

c\textbf{Discarding threshold.} The main computational cost in estimating Shapley value is computing the marginal contribution of each neuron. For a small subset of neurons ${Af}^{i}(O)$, our experiments find that after removing a small portion of neurons, ASR (attack success rate) of the networks will reduce to a low rate sharply. Thus, the marginal contribution of neurons after that can be negligible, and we can avoid calculating it, which saves substantial computational cost. Moreover, we mainly focus on top-$k$ neurons which are the most important in ASR when the network structure is complete and the performance is normal. Thus, we propose to discard neurons' marginal value after ASR is below a threshold, \eg 0.2. Note that we do not set the marginal value of the neurons to be zero after the model's performance reduces to a low rate. This is because if the neurons with larger Shapley value are in the latter part of the permutation, the marginal value of those neurons will be set to zero, making their Shapley value underestimated, especially when the number of average iterations is small. Our experiments also demonstrate that setting to zero can cause fluctuation and randomness in Shapley estimation, which needs a large average iteration to offset that negative effect.

\textbf{$\epsilon$-greedy acceleration.} Since we focus on neurons with top-$k$ largest Shapley values, neurons with larger Shapley value should be calculated more times to be estimated more precisely and get a more accurate sorting of them. However, because of discarding threshold, the neurons after ASR is under a threshold will be discarded and lose the opportunity to be computed. To improve their calculation times, we should assign neurons with the top Shapley values a higher probability to be at the front of the permutations. To this end, we propose an $\epsilon$-greedy based acceleration. 

$\epsilon$-greedy algorithm \cite{wunder2010classes}, as an optimization method, selects the best choice with probability $1-\epsilon$ and chooses from all the choices randomly with probability $\epsilon$. $\epsilon$-greedy algorithm is usually used in reinforcement algorithm and helps find the best choice in the action space \cite{hausknecht2015impact}. Thus, to improve estimation efficiency, we follow this idea and propose an $\epsilon$-greedy based algorithm, balancing exploration and utilization in finding the top-$k$ Shapley value neurons. We divide neurons into two groups based on the current Shapley value, estimated by the current average of each neurons' marginal values, top-$m$ and the other ($m \ge k$). We randomly permute neurons before the average iteration is smaller than $l$. Then after $l$, we utilize $\epsilon$-greedy, choosing a random neuron from top-$m$ with probability $1-\epsilon$ and choosing from the others with probability $\epsilon$ iteratively, and get a permutation to prune the neurons. 

\subsection{Trigger Reverse Synthesis}
We choose ASR as the metric function to estimate each neurons' Shapley value, and therefore, we need the backdoor dataset to calculate ASR. Furthermore, backdoor attacks manipulate DNNs to specific behaviors only on triggered images, indicating that the poisoned neurons are only activated on the backdoor images rather than normal images. Thus, reversing triggers from the backdoor networks, as an important part, can help the removal of backdoor neurons in poisoned models. Intuitively, backdoor attacks make use of DNNs' overfitting feature to create a shortcut for the trigger to cause DNNs' misclassification. We can use the trigger reverse synthesis to reverse backdoor trigger in the models and generate the reversed backdoor dataset\cite{wang2019neural}. 

We first inject class-specific reversed trigger ${T}_{c}$ into clean images and get the triggered image $a_{c}$ as follows:
\begin{equation}
a_{c}=\left(1-M_c\right) \odot a+{M}_{c} \odot {T}_{c}
\end{equation}
where ${M}_{c}$ represents the mask for class c, deciding the location and intensity of trigger being injected into original images, $a$ represents the original image, $ {T}_{c}$ represents the trigger patter for class c and $\odot$ means Hadamard product. Similar to adversarial example generation, we optimize on networks' misclassification and trigger size to reverse backdoor. We use Cross-Entropy loss to optimize misclassification of triggered images to class $c$ and $L1$ norm of the mask to optimize the trigger size. We sum the above objectives and get the equation as follows:

\begin{equation}
 \min _{M_c, T_c} \quad CE(y_{c}, f(a_c))+\lambda \cdot|M_c|_{1} \quad for \quad a \in A 
\end{equation}
where $y_{c}$ represents label for class $c$, $A$ represents clean images available, $CE(\cdot)$ represents Cross-Entropy loss, $|M_c|_{1}$ represents $L1$ norm of the mask, and $\lambda$ represents the trade-off parameter.

From the above method, we can get the reversed trigger for each target class. However, judging whether the network is poisoned and which is the target label is still a problem. Intuitively, since backdoor training produces a shortcut for the backdoor trigger in the poisoned models, the reversed trigger for the target label is the smallest among all the classes. Thus, we can get the reversed trigger and target label by finding the smallest trigger in trigger reverse synthesis.

\textbf{Backdoor model detection.} First, the L1 norm of the reversed trigger for the target label is much smaller than the others. Thus, L1 norm for the target label can be seen as an outlier from the other triggers, and we can use anonamly detection method to find the target label. We employ MAD (Median Absolute Deviation) to judge whether the models are poisoned. By MAD, supposing that mask norms obey normal distribution \cite{rousseeuw1993alternatives}, any anomaly index $I=d_{i}/MAD$ larger than a specific value will be treated as an outlier, where $d_{i}$ is the absolute deviation between triggers' L1 norm and their median.

However, in experiments, we find that the reversed triggers for some classes can't converge to a small L1 norm, and their norms are abnormally larger than expected, causing a false positive in backdoor detection. Different from \cite{wang2019neural}, since we only focus on whether the smallest reversed trigger is an outlier, we can just apply MAD to the set of the triggers whose norms are smaller than the median to avoid anomaly large norms. We define their deviations' set as $D_{small}= \lbrace d_1,\ldots,d_l \rbrace$. Furthermore, because normal distribution is symmetrical, the median of $D_{small}$ can be used to replace the median of all labels' deviation as MAD. Then, we can use MAD to estimate the standard deviation $\sigma$ of the distribution of norms and use it to detect backdoor model with confidence probability $p$, expressed as follows:

\begin{equation}
\sigma = \frac{MAD}{\Phi^{-1}(\frac{3}{4})} \approx 1.4826\cdot MAD  
\end{equation}

%P(\frac{d_{small}}{\sigma}\leq d)=2\Phi(d)-1=p
\begin{equation}
\frac{d_{i}}{\sigma} \leq d=\Phi^{-1}(\frac{p+1}{2}),\quad d_{i}=d_{1},d_{2},\cdots, d_{l}
\end{equation}
where $\Phi(\cdot)$ represents cumulative probability distribution of standard normal distribution and $d$ represents max bound for $d_{i}/\sigma$ with probability $p$. The norm with the deviation larger than $\sigma \cdot \Phi^{-1}(\frac{p+1}{2})$ is an outlier and the model has been poisoned. 
\subsection{Data-free Backdoor Mitigation}

As we mentioned above, we can mitigate backdoor from models with few images using ShapPruning. Then we further research on a no clean data situation and propose a data-free ShapPruning method. To help data-free backdoor mitigation with Shapley estimation, we need to reverse training images from the poisoned models first. Recent works in transfer learning show that the batch normalization layer (BN layer) can be used to recover better images from the trained models and improve transfer efficiency \cite{choi2020data,haroush2020knowledge,yin2020dreaming}. Furthermore, because backdoor attacks only poison a very small portion of training datasets, they won't affect the information in BN layers, and thus, we can use BN layers to better reverse images. Then we can express the difference of the mean and variance between the recovered data and original training data in the model's each BN layer as follows:
\begin{equation}
L_{bn}(x) = \sum_{i}{div(N(\mu_{i}(x),\sigma_{i}(x))),N(\mu_{i},\sigma_{i}))}
\end{equation}
where $div(\cdot)$ represents divergence, $N(\mu_{i}(x),\sigma_{i}(x))$ represents mean and variance of recovered data $x$ on BN layer $i$, and $N(\mu_{i},\sigma_{i})$ represents mean and variance recorded in BN layer $i$. Furthermore, considering the image prior information,  we got the prior loss as follows:
\begin{equation}
L_{pr}(x) = \alpha_{1}L_{V}(x) + \alpha_{2}L_{norm}(x)\\
\end{equation}
where $L_{V}(x)$ represents the variation of images, $L_{norm}(x)$ represents images' norm, and $\alpha_{1},\alpha_{2}$ are hyper-parameters. Then, with the analysis above, we use the total loss $L_{total}$ to reconstruct the training images from poisoned models for estimating Shapley value:
\begin{gather}
L_{total}(x) = \alpha CE(f(x),y) +\beta L_{bn}(x) + \gamma L_{pr}(x)
\end{gather}
where $CE(\cdot)$ represents the Cross Entropy loss, $f(\cdot)$ represents the trained model, $y$ represents the target label to reconstruct images and $\alpha,\beta,\gamma$ are hyper-parameters. 

\textbf{Mixture-mode.} Furthermore, because there is still a difference between clean images and recovered images, our experiments find that there is a larger accuracy degradation during data-free ShapPruning. Therefore, we propose a mixture-mode and try to combine information of Acc (accuracy) and ASR. We calculate Shapley value of Acc and ASR separately and find neurons with top-$k$ ASR Shapley value and bottom-$l$ Acc Shapley value to prune. And our experiments demonstrate that this way can help us locate the neurons which are only important to backdoor but not overall accuracy, locating poisoned neurons more accurately.

\subsection{Pruning Based on Estimated Shapley Value}
Finally, we summarize our framework illustrated in \cref{shapley_pruning}. We first use trigger reverse synthesis to get the reversed trigger and the target label. Then we inject the trigger into clean data (in the data-free situation, clean data is recovered from poisoned models), and get the reversed backdoor dataset. Then, using ASR as a measurement, we implement the accelerated Shapley value estimation method to get top-$k$ neurons with the largest Shapley value. Finally, we prune the target network with the top neurons, fine-tune the network with the clean data available, and offer backdoor-free networks to the users.

\section{Experiment}
We evaluate our backdoor defense method with five mainstream tasks against five common attacks, BadNets Attack \cite{gu2017badnets}, Trojan Attack \cite{liutrojaning}, Physical Key Attack \cite{chen2017targeted}, Input-Aware Attack \cite{nguyen2020input} and WaNet Attack\cite{nguyen2020wanet} in data-insufficient situations on VGG \cite{simonyan2014very} and ResNet \cite{he2016deep}, and design a series of experiments to test its effectiveness and robustness.

\subsection{Experimental Setup}
We compare ShapPruning with four existing methods Fine Pruning (FP) \cite{liu2018fine}, Neural Cleanse (NC) \cite{wang2019neural}, GangSweep (GS) \cite{zhu2020gangsweep} and DeepInspect (DI) \cite{chen2019deepinspect} on the following five datasets (1).MNIST (2).CIFAR10 (3).CIFAR100 (4).GTSRB (5).YouTubeFace.

\begin{figure}
  \centering
  \begin{subfigure}{0.23\linewidth}
  \centering
  \includegraphics[width=1.5cm]{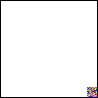}
    \caption{}
  \end{subfigure}
  \hfill
  \begin{subfigure}{0.23\linewidth}
  \centering
    \includegraphics[width=1.5cm]{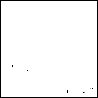}
    \caption{}
  \end{subfigure}
  \begin{subfigure}{0.23\linewidth}
  \centering
  \includegraphics[width=1.5cm]{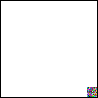}
    \caption{}
  \end{subfigure}
  \hfill
  \begin{subfigure}{0.23\linewidth}
  \centering
    \includegraphics[width=1.5cm]{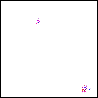}
    \caption{}
  \end{subfigure}
  \begin{subfigure}{0.23\linewidth}
  \centering
  \includegraphics[width=1.5cm]{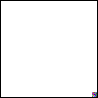}
    \caption{}
  \end{subfigure}
  \hfill
  \begin{subfigure}{0.23\linewidth}
  \centering
    \includegraphics[width=1.5cm]{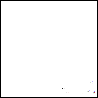}
    \caption{}
  \end{subfigure}
  \begin{subfigure}{0.23\linewidth}
  \centering
  \includegraphics[width=1.5cm]{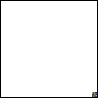}
    \caption{}
  \end{subfigure}
  \hfill
  \begin{subfigure}{0.23\linewidth}
  \centering
    \includegraphics[width=1.5cm]{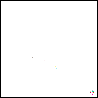}
    \caption{}
  \end{subfigure}
 \caption{Original and reversed triggers in BadNets. (a), (c), (e), (g) are original triggers for CIFAR10, YouTubeFace, MNIST, GTSRB with the size of $10\times10$, $10\times10$, $5\times5$, $5\times5$ respectively. (b), (d), (f), (h) are reversed triggers for CIFAR10, YouTubeFace, MNIST, GTSRB using trigger reverse synthesis.}
\label{reverse_trigger}
\end{figure}

\textbf{Attack configurations for BadNets.} In our experiments, BadNets poisons images with a random colored square shown in \cref{reverse_trigger}. The trigger sizes are $5\times5$ or $10\times10$ to test our defense's robustness against different trigger sizes. We resize the images to $96\times96$, and thus, the trigger size is about $1\%$ of the image size and injection ratio is $1\%$. Our experiments are based on a VGG11-based model which is usually used in model compression tasks.

\textbf{Attack configurations for Trojan Attack.} We generate the trojan trigger shown in \cref{physical attack} with an initial square mask using gradient descent. Then, with the generated trigger, we fine-tune the trained model on the linear layers to inject backdoor into pre-trained models.   

\textbf{Attack configurations for Physical Key Attack.} Physical Key Attack uses a pair of commodity glasses shown in \cref{physical attack} rather than a small square to inject backdoor into the model and can be more imperceptible. 

\textbf{Attack configurations for Input-Aware Attack.} It uses a generator to generate sample-specific triggers. We set the Input-Aware Attack in all-to-one mode and attain a model with $99.41\%$ ASR with the same setting in \cite{nguyen2020input}. 

\textbf{Attack configurations for WaNet Attack.} It uses warping-based triggers to generate sample-specific, undetectable triggers. We defend against WaNet based on the same setting in \cite{nguyen2020wanet}.

\textbf{Available data.} We suppose the defender can only get a small amount of clean data, specifically, one image per class in the few-shot setting \eg only $10$ images available for MNIST and CIFAR10. Furthermore, we propose a more strict condition in \cref{data-free} where only the poisoned models are available but no clean data to help mitigate the backdoor.

\subsection{Shapley Pruning}
In this subsection, we compare ShapPruning with other defense methods and prove its effectiveness.

\begin{figure}
  \centering
  \begin{subfigure}{0.22\linewidth}
  \centering
  \includegraphics[width=0.90\linewidth]{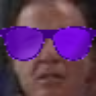}
    \caption{}
  \end{subfigure}
  \hfill
  \begin{subfigure}{0.22\linewidth}
  \centering
    \includegraphics[width=0.90\linewidth]{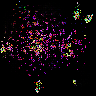}
    \caption{}
  \end{subfigure}
   \begin{subfigure}{0.22\linewidth}
  \centering
  \includegraphics[width=0.90\linewidth]{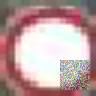}
    \caption{}
  \end{subfigure}
  \hfill
  \begin{subfigure}{0.22\linewidth}
  \centering
    \includegraphics[width=0.90\linewidth]{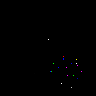}
    \caption{}
  \end{subfigure}
  
 \caption{Trigger synthesis for Physical Key Attack and Trojan Attack. (a), (c) are examples of poisoned data for Physical Key Attack and Trojan Attack, and (b), (d) are the reversed trigger generated by trigger synthesis for Physical Key and Trojan Attack.}
\label{physical attack}
\end{figure}

\textbf{Trigger reverse synthesis.} We first use trigger reverse synthesis to get the reversed trigger in \cref{reverse_trigger,physical attack} where we find the reversed triggers and original triggers are in a similar location of the images, but have a relative difference in shape and color, caused by the insufficiency of data. Also, trigger reverse synthesis penalizes L1 norm, causing reversed triggers smaller than the original ones. These mismatches lead to some performance degradation in the backdoor mitigation compared with defense with the original trigger. However, despite the differences, ShapPruning can still locate the poisoned neurons precisely and mitigate backdoor with different trigger sizes.

\begin{figure*}
  \centering
  \begin{subfigure}{0.24\linewidth}
  \centering
  \includegraphics[width=0.8\linewidth]{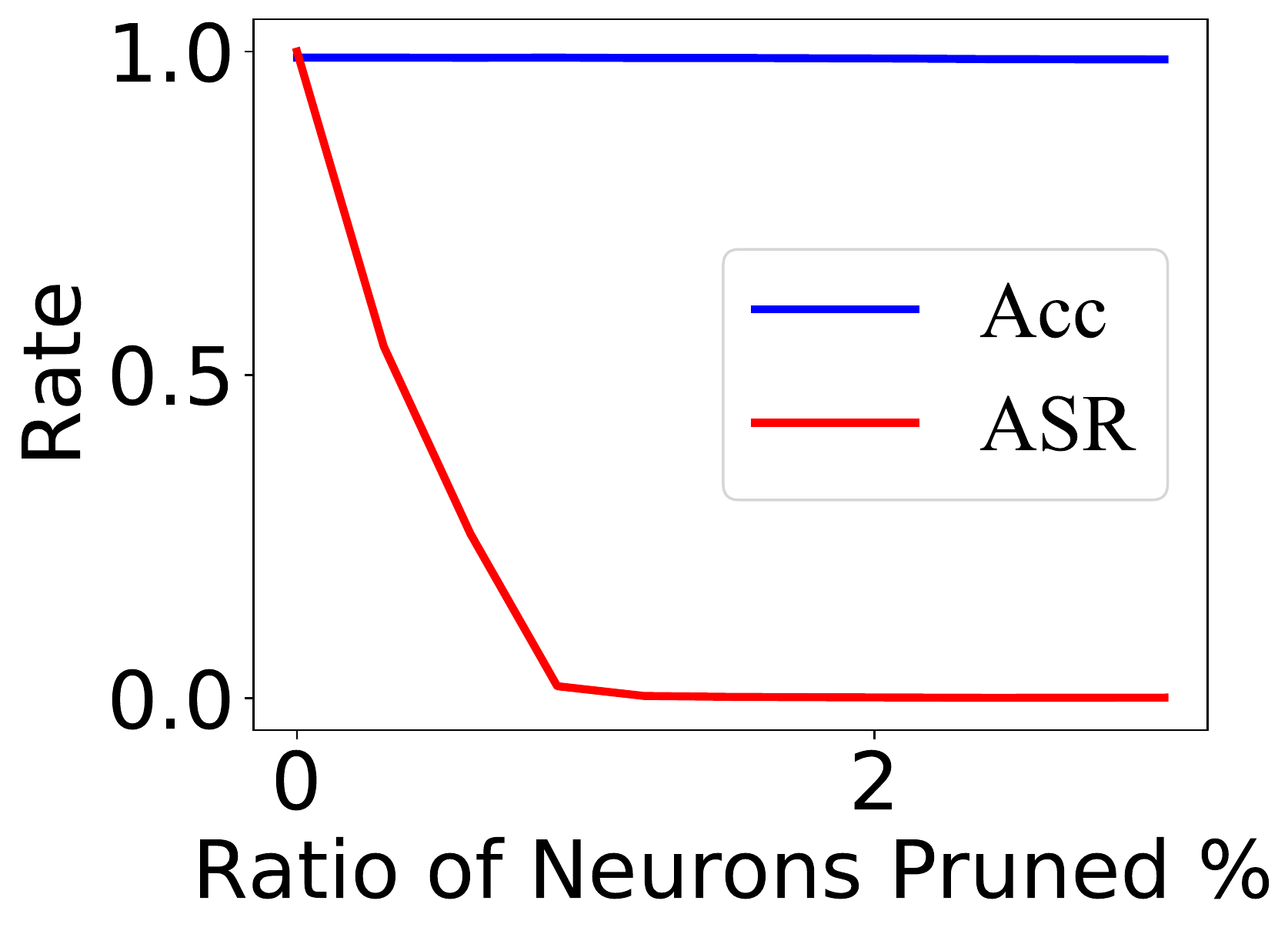}
    \caption{ShapPruning on MNIST}
  \end{subfigure}
  \hfill
  \begin{subfigure}{0.24\linewidth}
  \centering
    \includegraphics[width=0.8\linewidth]{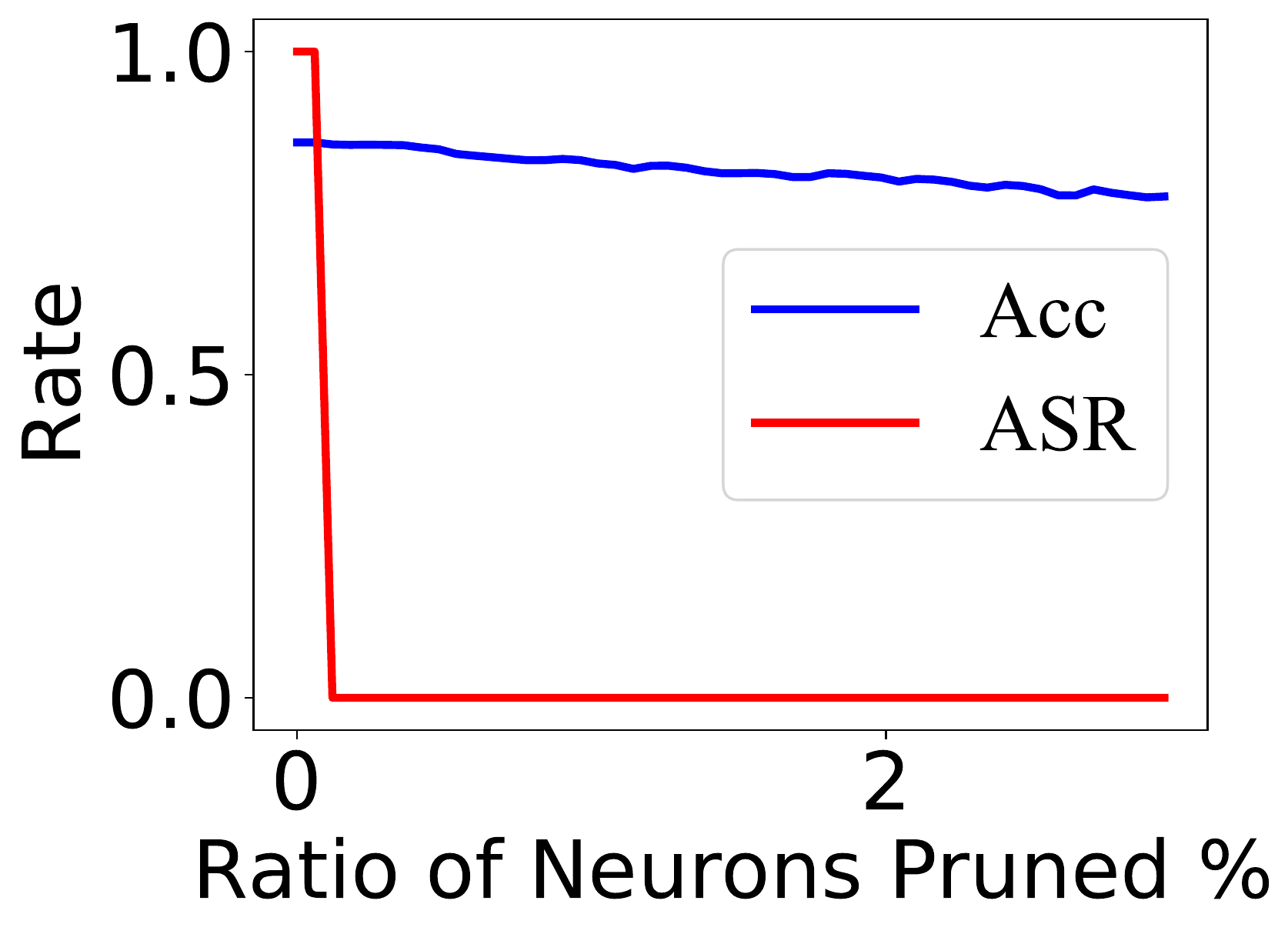}
    \caption{ShapPruning on CIFAR10}
  \end{subfigure}
  \begin{subfigure}{0.24\linewidth}
  \centering
  \includegraphics[width=0.8\linewidth]{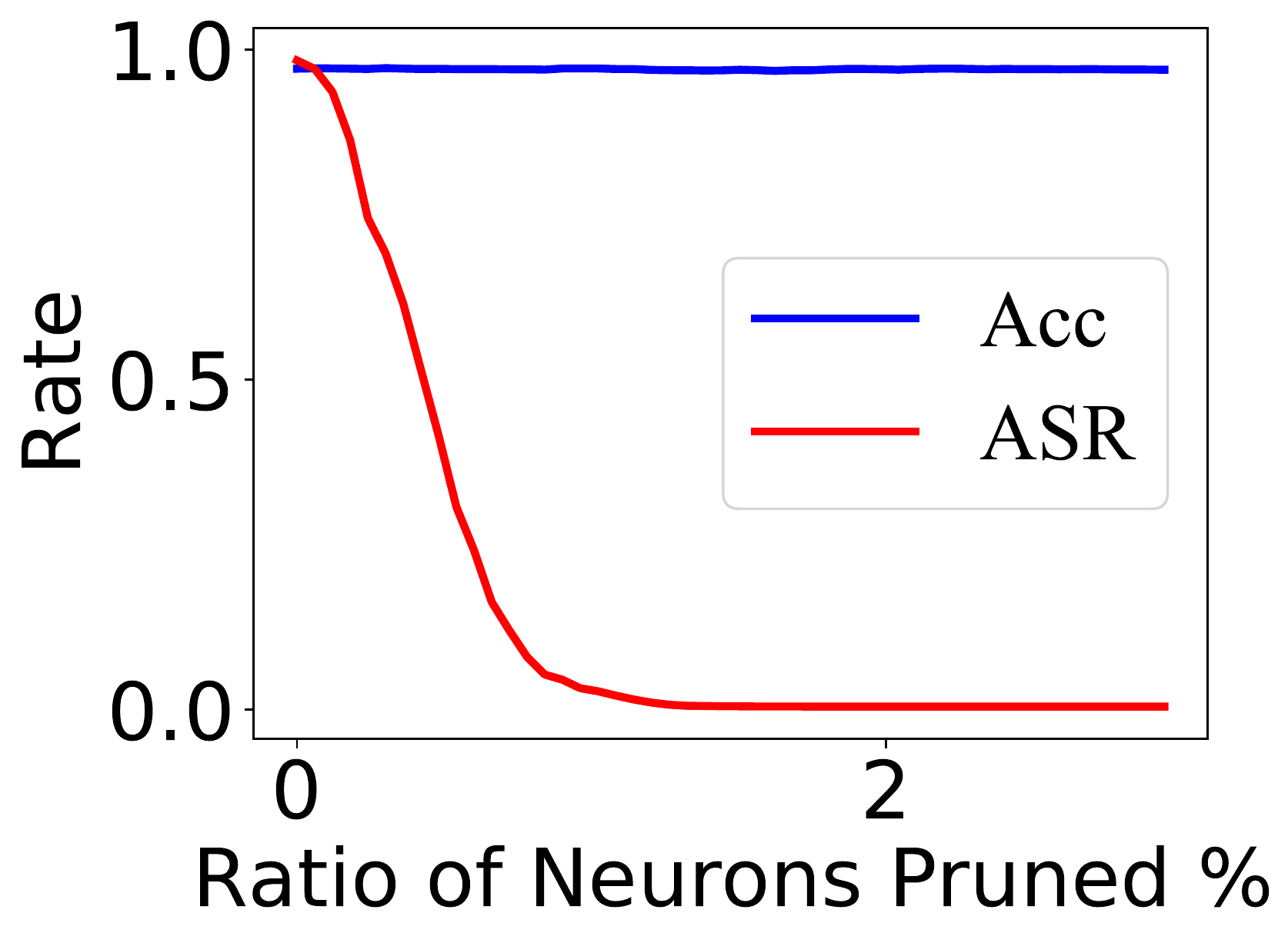}
    \caption{ShapPruning on GTSRB}
  \end{subfigure}
  \hfill
  \begin{subfigure}{0.24\linewidth}
  \centering
    \includegraphics[width=0.8\linewidth]{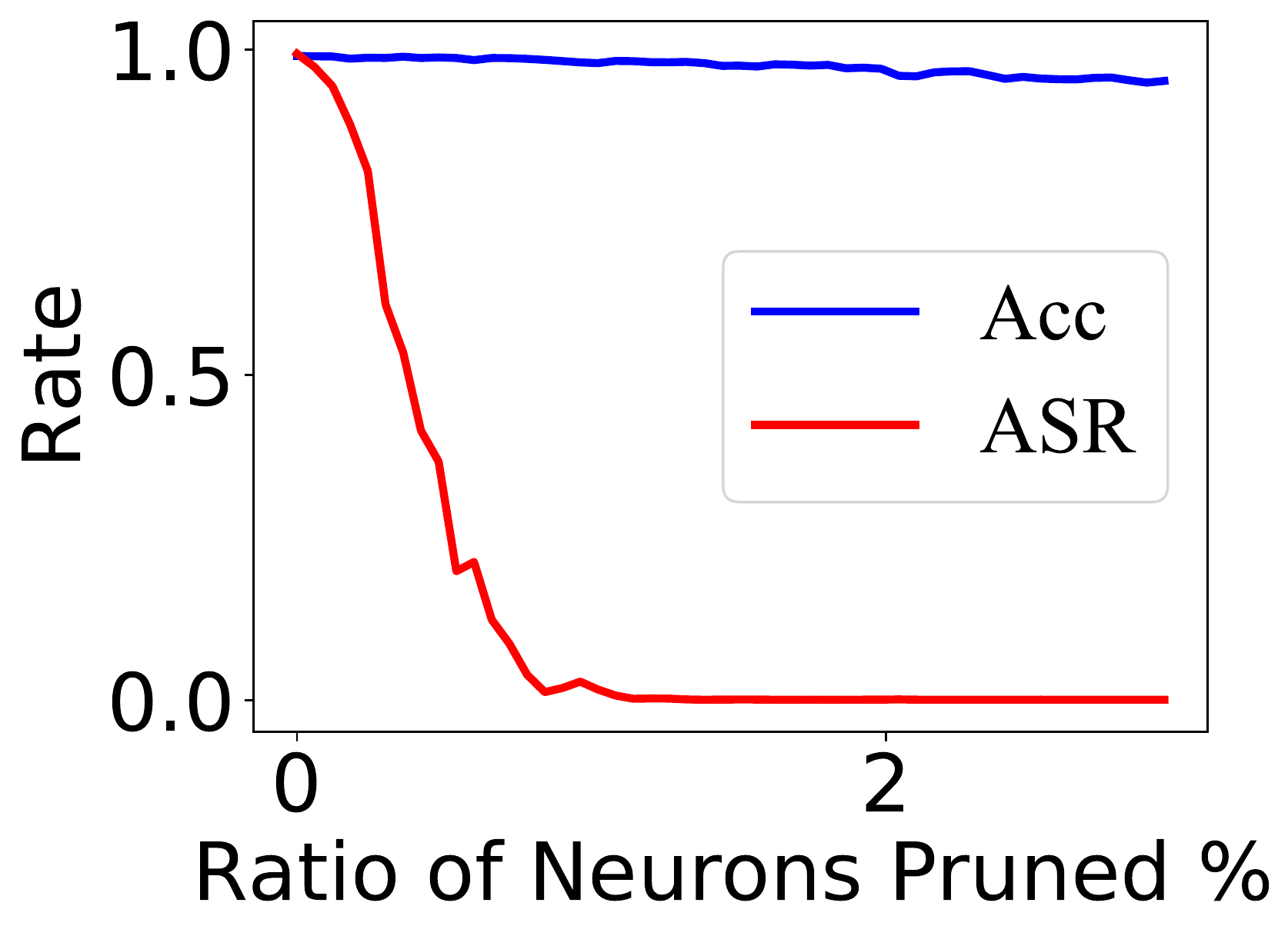}
    \caption{ShapPruning on YouTubeFace}
  \end{subfigure}
  \begin{subfigure}{0.24\linewidth}
  \centering
  \includegraphics[width=0.8\linewidth]{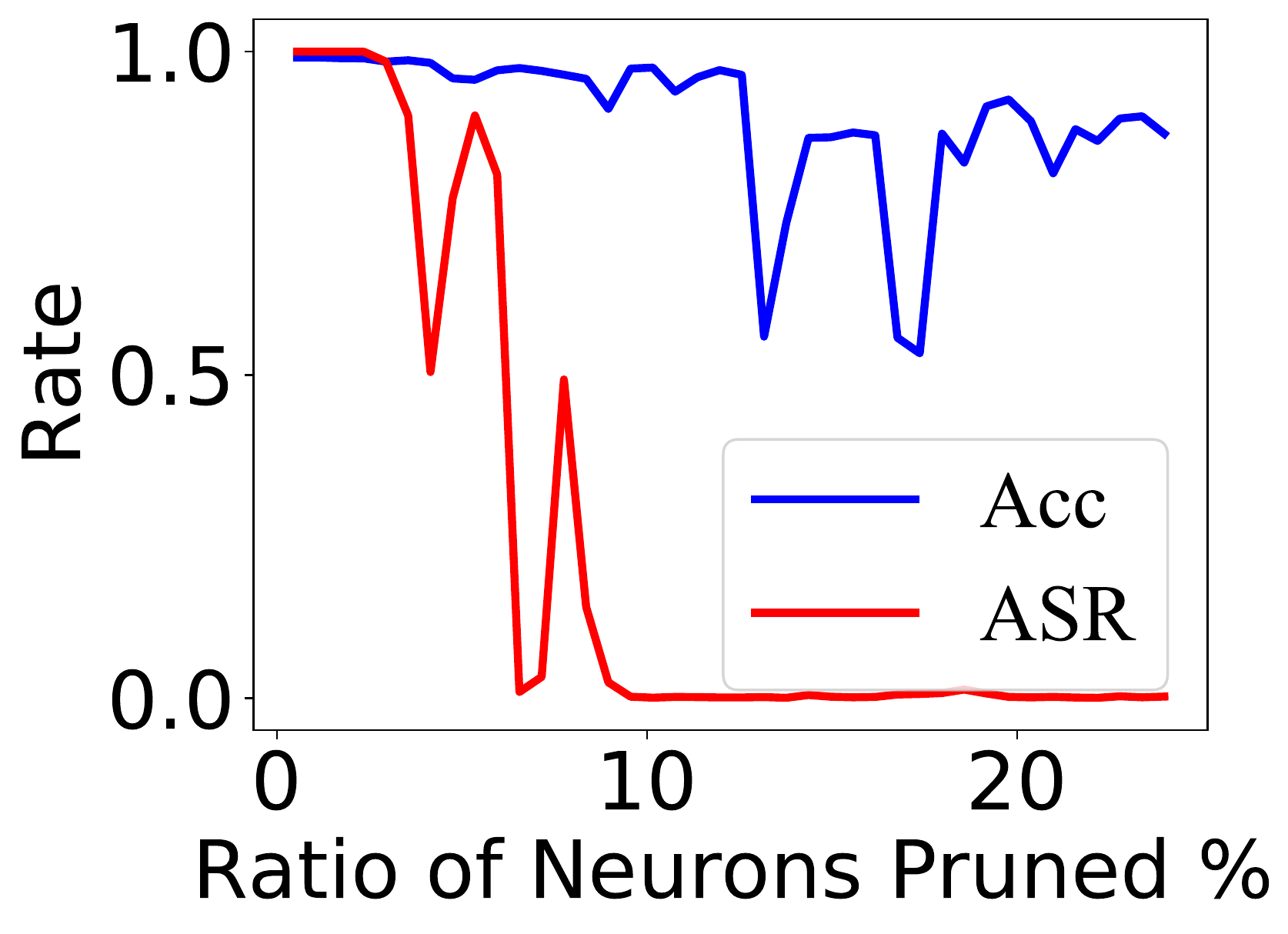}
    \caption{Fine Pruning on MNIST}
  \end{subfigure}
  \hfill
  \begin{subfigure}{0.24\linewidth}
  \centering
    \includegraphics[width=0.8\linewidth]{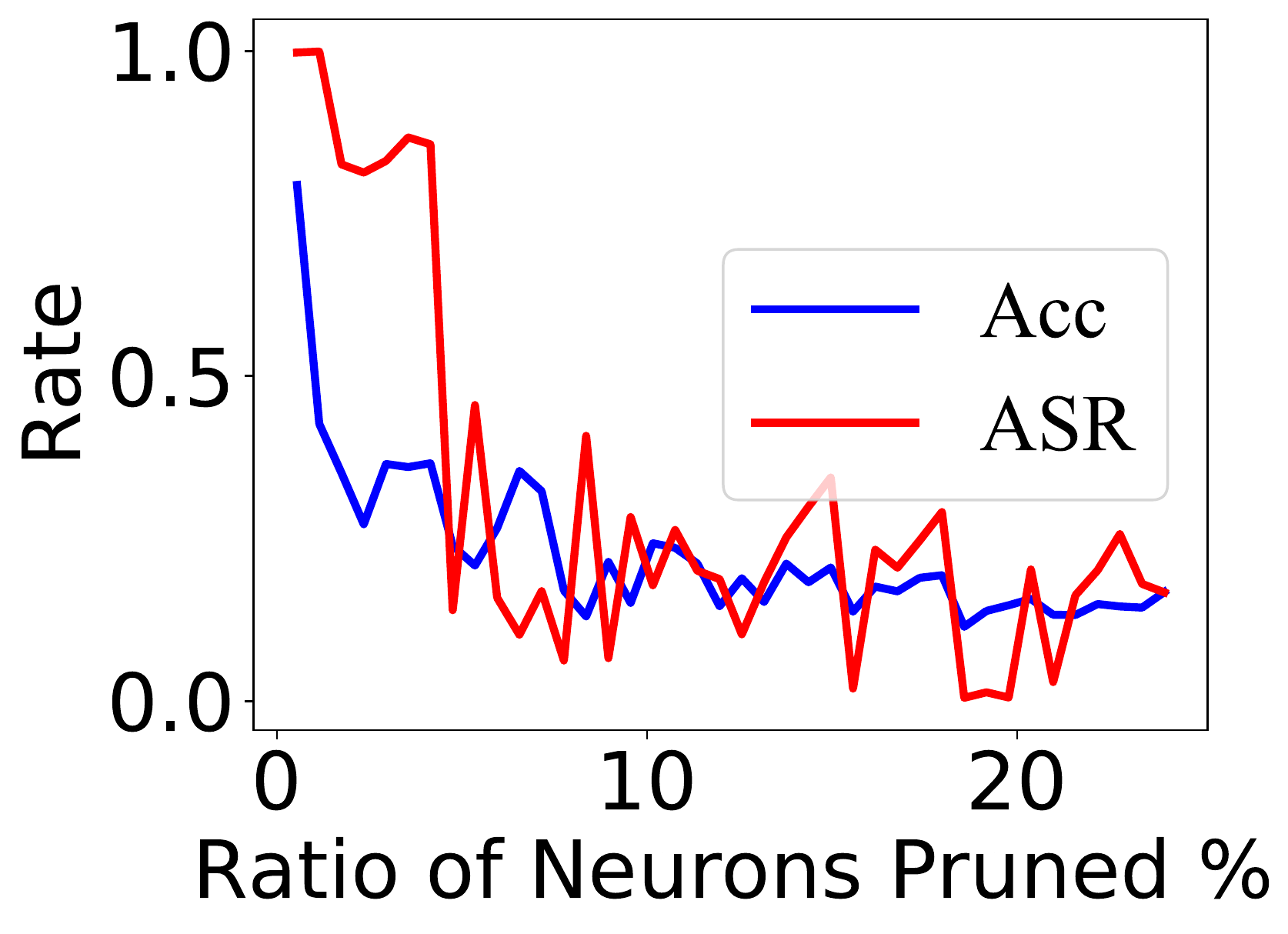}
    \caption{Fine Pruning on CIFAR10}
  \end{subfigure}
  \begin{subfigure}{0.24\linewidth}
  \centering
  \includegraphics[width=0.8\linewidth]{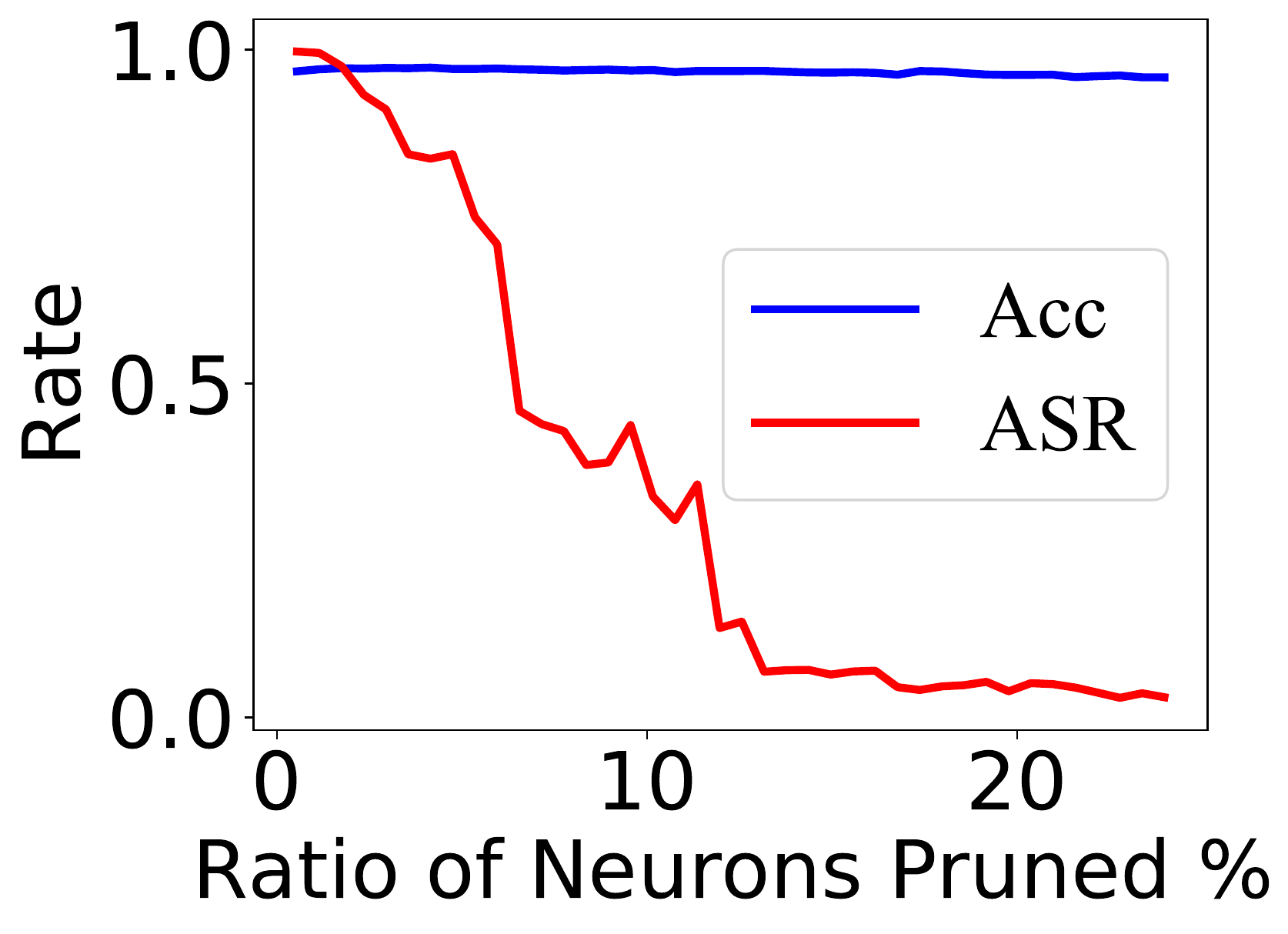}
    \caption{Fine Pruning on GTSRB}
  \end{subfigure}
  \hfill
  \begin{subfigure}{0.24\linewidth}
  \centering
    \includegraphics[width=0.8\linewidth]{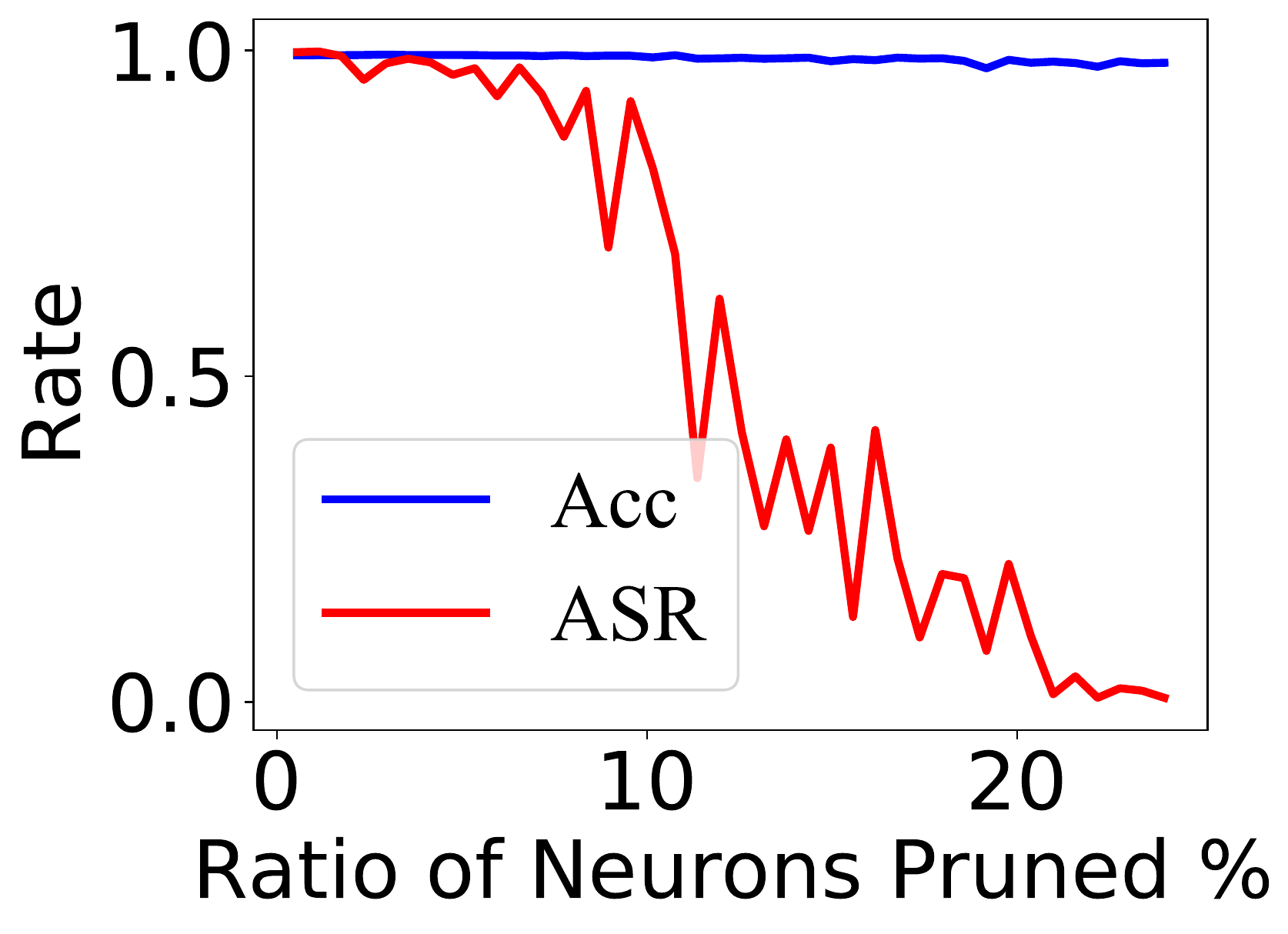}
    \caption{Fine Pruning on YouTubeFace}
  \end{subfigure}
 \caption{Acc and ASR fluctuation when pruning neurons guided by ShapPruning or Fine Pruning. (a)-(d) are for ShapPruning with max $3\%$ of all the neurons pruned and (e)-(h) are for Fine Pruning with max $25\%$ of all the neurons pruned.}
\label{Fine and Shapley}
\end{figure*}

\textbf{Pruning based on Shapley value.} With the reversed poisoned data, we prune neurons in the order produced by $\epsilon$-greedy and compute ASR decline in output iteratively, finding $50$ average iterations are precise enough for estimating Shapley value and locating poisoned neurons. We compare our method with other three common methods including Fine Pruning (FP), Neural Cleanse (NC), and GangSweep (GS). From \cref{tab:performance_comparison}, ShapPruning mitigates backdoor best in poisoned models at the price of a tiny accuracy decline with only one image per class. On the contrary, other methods can't clean backdoor attacks with that few images. Especially, all the other defenses perform weaker in MNIST and CIFAR10 with only $10$ clean images, which are fewer than the other two datasets, GTSRB and YouTubeFace. 

We suppose the poor performance of Neural Cleanse (NC) and GangSweep (GS) is caused by both the gap between the original and reversed trigger, and the weak generalization of training with a small amount of clean data. We explain the trigger gap's influence on mitigation degradation with the concepts from adversarial training. Neural Cleanse or GangSweep is similar to adversarial training \cite{zhang2019defense,shafahi2020universal}, which meets with performance degradation and poor generalization against different attacks. Thus, NC or GS, similar to adversarial training, behaves poorly with bigger trigger gaps, which is also found in \cite{pang2020trojanzoo}. Furthermore, we show Acc and ASR fluctuation during ShapPruning and Fine Pruning in \cref{Fine and Shapley}. It demonstrates that ShapPruning can remove backdoor with only $1\%$ of total neurons, compared with about $25\%$ neurons removal in Fine Pruning. Fine Pruning, with that number of neurons pruned, may cause network structure changes and accuracy decline. Also, the insufficiency of clean data can weaken fine-tuning process and cause large accuracy fluctuation, especially in MNIST and CIFAR10 with only 10 images.

\begin{table*}[h]
  \centering  
  \setlength{\tabcolsep}{2mm}{
    \begin{tabular}{c c c c c c c c c c c c c}  
    \hline  
    {Benchmark}&  
    \multicolumn{2}{c}{Before}&\multicolumn{2}{c}{FP\cite{liu2018fine}} &\multicolumn{2}{c}{NC\cite{wang2019neural}} &\multicolumn{2}{c}{GS\cite{zhu2020gangsweep}}  &\multicolumn{2}{c}{ShapPruning}
    &\multicolumn{2}{c}{ShapPruning/o}
    \cr\cline{2-13}  
    ($\%$)&Acc&ASR&Acc$\uparrow$&ASR$\downarrow$&Acc$\uparrow$&ASR$\downarrow$&Acc$\uparrow$&ASR$\downarrow$&Acc$\uparrow$&ASR$\downarrow$&Acc$\uparrow$&ASR$\downarrow$\cr 
    \hline  
    MNIST & $99.02$  & $100.00$ & $97.00$ & $3.02$ & $98.64$ & $29.87$ & $95.30$ & $80.33$ &$98.99$ & $0.34$ &$99.06$ & $0.56$ \cr
    CIFAR10 & $86.05$ & $99.57$ & $35.39$ & $10.19$ & $78.98$ & $46.32$  & $83.45$ & $100.00$& $85.63$ & $0.06$ &$85.66$ & $0.03$ \cr 
    GTSRB & $97.03$  & $99.60$ & $96.26$ & $6.16$ & $96.69$ & $4.76$ &  $96.63$ & $1.11$ &  $96.94$ & $0.49$ &  $97.16$ & $0.46$ \cr
    YouTubeFace & $98.93$ & $99.82$ & $97.49$ & $0.61$ & $95.66$ & $7.38$ & $90.90$ & $0.58$ &$98.61$ & $0.35$ &$98.67$ & $0.34$  \cr\hline
    Input-Aware & $99.41$ & $99.37$ & $98.12$ & $2.66$ & $99.32$ & $43.55$ & $88.85$ & $32.05$ &$99.29$ & $0.15$ &$99.35$ & $0.24$\cr
    Trojan Attack& $97.08$ & $92.06$ & $16.32$ & $2.56$ & $95.01$ & $2.01$ & $96.33$ & $10.91$ &$96.03$ & $0.98$ &$96.44$ & $0.64$\cr
    Physical Key & $98.39$ & $100.00$ & $90.70$ & $0.05$ & $98.49$ & $64.34$   & $97.21$ & $54.21$ & $95.94$ & $0.60$ & $97.26$ & $0.08$ \cr
    WaNet & $98.21$ & $98.10$ & $37.90$ & $10.82$ & $97.92$ & $97.11$   & $96.28$ & $90.24$ & $97.54$ & $0.93$ & $97.73$ & $0.32$ \cr\hline
    ResNet-18 & $95.17$ & $100.00$ & $17.03$ & $0.79$ & $90.44$ & $43.10$ & $89.46$ & $57.73$ & $92.25$ & $0.48$ & $92.71$ & $0.20$\cr
    ResNet-34  & $98.37$ & $99.98$  & $97.93$ & $0.19$ & $98.65$ & $0.25$ & $56.84$ & $6.55$& $98.49$ & $0.07$& $98.51$ & $0.05$ \cr\hline
    
    \end{tabular}  
    }
    \caption{Different defenses methods against five common attacks and two common architectures (VGG and ResNet), where ShapPruning/o represents ShapPruning with original trigger. The first four lines show defenses against BadNets on four common datasets, the fifth to eithth lines show defenses against four different attacks (Input-Aware Attack on MNIST, Trojan Attack on GTSRB, Physical Key on YouTubeFace and WaNet on GTSRB), and the ninth and tenth lines show defenses against ResNet (ResNet-18 on GTSRB and ResNet-34 on YouTubeFace). We record their Acc (higher is better) and ASR (lower is better) in the table.}
    \label{tab:performance_comparison}  
\end{table*}

\begin{figure*}
  \centering
  \begin{subfigure}{0.33\linewidth}
  \centering
  \includegraphics[width=0.8\linewidth]{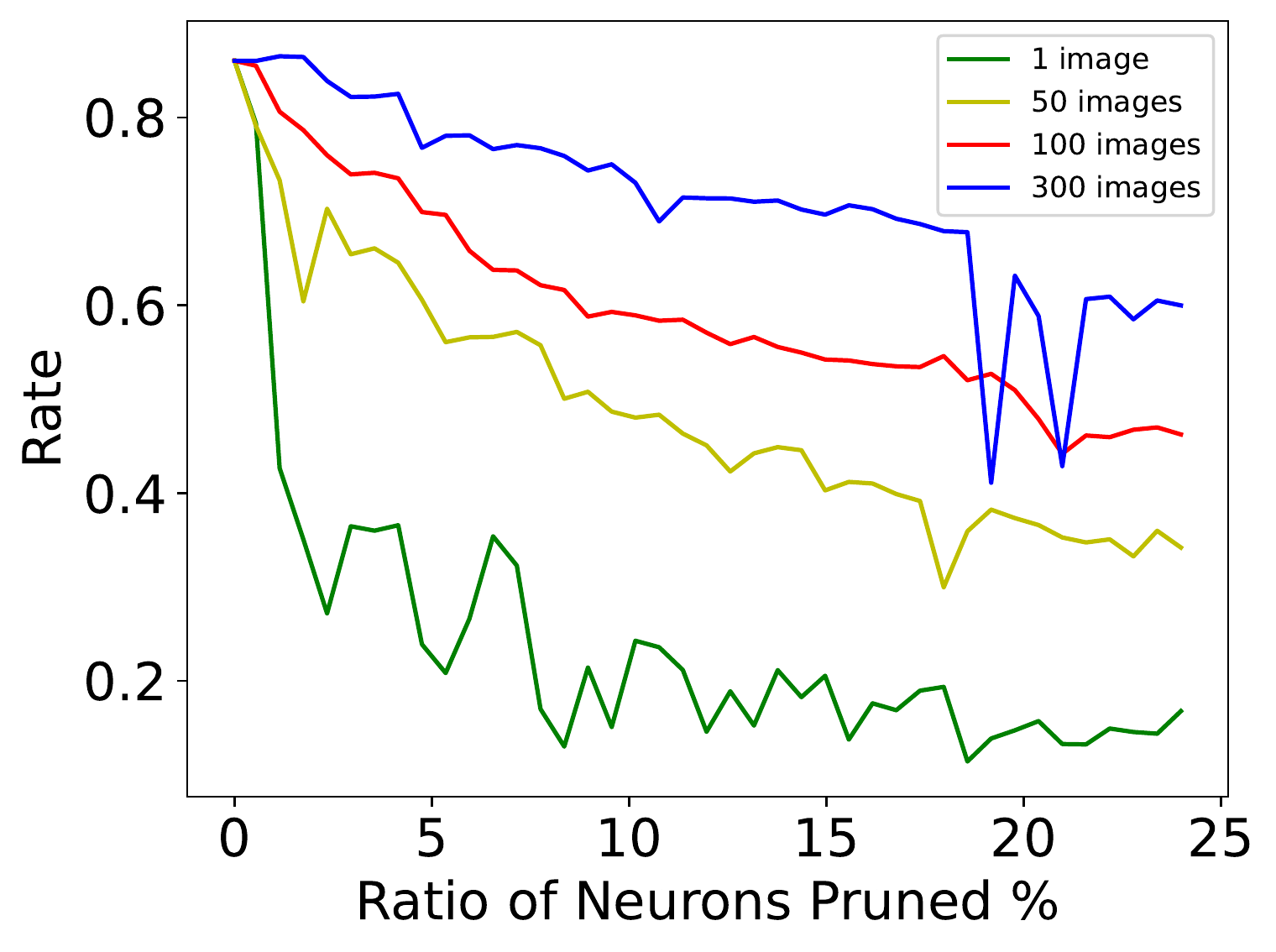}
    \caption{Acc for Fine Pruning}
  \end{subfigure}
  \hfill
  \begin{subfigure}{0.33\linewidth}
  \centering
    \includegraphics[width=0.8\linewidth]{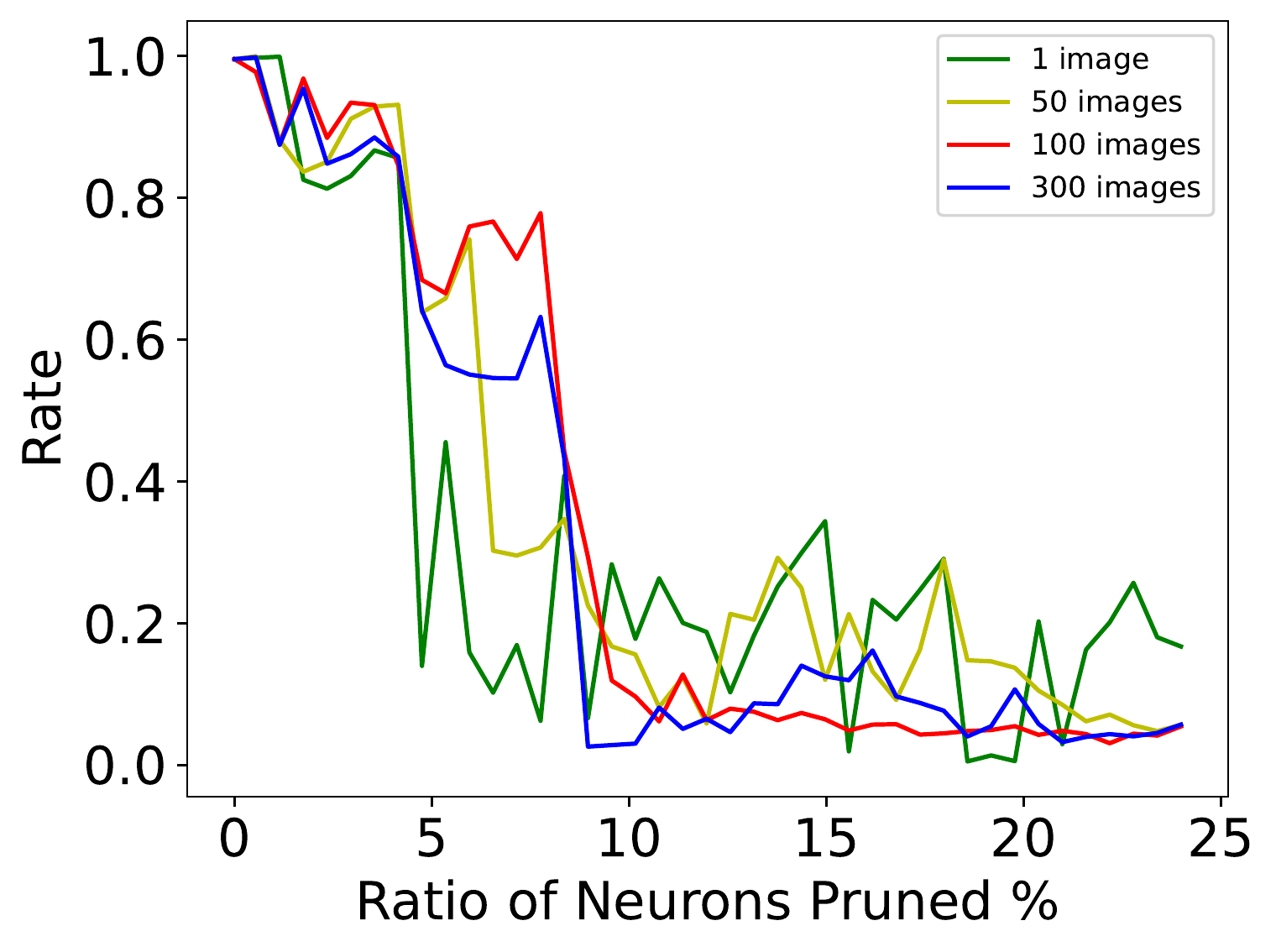}
    \caption{ASR for Fine Pruning}
  \end{subfigure}
  \begin{subfigure}{0.33\linewidth}
  \centering
  \includegraphics[width=0.8\linewidth]{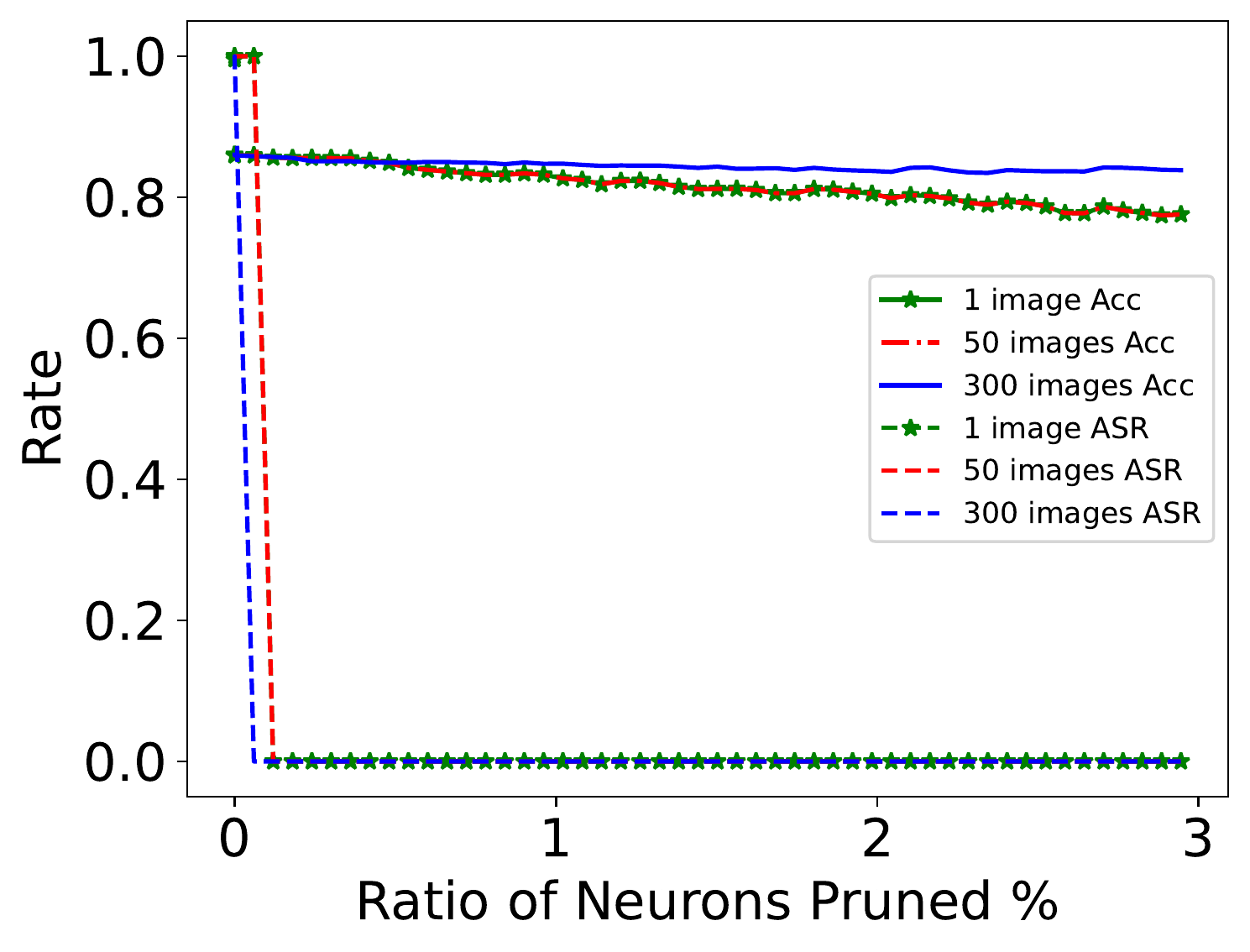}
    \caption{Acc and ASR for Shapley Pruning}
  \end{subfigure}
 \caption{Fine Pruning and ShapPruning with different sizes of datasets on CIFAR10. We test on 4 datasets with different amounts of clean data with 1 image, 50 images, 100 images, 300 images per class respectively for Fine Pruning and 1 image, 50 images, 300 images per class respectively for ShapPruning.}
\label{pruning on different dataset}
\end{figure*}

\begin{table*}[]
  \centering  
    \begin{tabular}{c c c c c c c c c c c }  
    \hline  
    {Benchmark}&  
    \multicolumn{2}{c}{Before}&\multicolumn{2}{c}{FP\cite{liu2018fine}} &\multicolumn{2}{c}{NC\cite{wang2019neural}} &\multicolumn{2}{c}{DI\cite{chen2019deepinspect}}  &\multicolumn{2}{c}{ShapPruning } 
    \cr\cline{2-11}  
    $(\%)$&Acc&ASR&Acc$\uparrow$&ASR$\downarrow$&Acc$\uparrow$&ASR$\downarrow$&Acc$\uparrow$&ASR$\downarrow$&Acc$\uparrow$&ASR$\downarrow$\cr 
    \hline  
    CIFAR10 & $86.51$  & $100.00$ & $27.27$ & $0.00$ & $82.48$ & $56.71$ &$48.35$ & $30.26$ & $83.04$ &$0.90$ \cr
    CIFAR100 & $62.08$ & $99.98$ & $35.39$ & $10.19$ & $47.12$ & $27.82$ & $53.76$ & $81.44$ & $59.51$ & $0.89$\cr 
    Trojan Attack & $97.08$  & $92.06$ & $22.49$ & $71.32$ & $95.17$ & $2.82$ &  $84.43$ & $33.24$ &$96.21$ & $0.14$ \cr\hline
    
    \end{tabular}  
    \caption{Different defense methods against different attacks and architectures in the data-free situation.}
    \label{tab:data-free}  
\end{table*}

 \textbf{Defense against different attacks.} We also defend against different attacks to test our method's robustness. We first show reversed triggers in \cref{physical attack}, where an obvious reversed gap is found. Based on the reversed triggers, we use different defense methods to mitigate backdoor, shown in \cref{tab:performance_comparison}. Also, we defend against Input-Aware Attack and WaNet Attack which both inject sample-specific triggers into images to activate the backdoor. Our experiment demonstrates that although there are different triggers for different samples, sample-specific attacks still rely on a small number of sensitive neurons to activate backdoor and our method can find them precisely.

\textbf{Time consumption.} We conducted our experiments on Titan RTX GPU with 24GB memory and recorded time consumption for different methods in mitigating backdoor attacks. We compare our method with Neural Cleanse in GTSRB and find our method only consumes 585.95 seconds with 50 average iterations' Shapley estimation to get results in \cref{tab:performance_comparison} after 671.13 seconds spent on trigger reverse. On the contrary, Neural Cleanse consumes 704.54 seconds which is 1.7x faster than our method. However, Neural Cleanse needs much more data and can't completely remove triggers in the few-shot setting. Furthermore, our method is time-saving and needs much less clean data compared with training a clean model from scratch.

\subsection{Defense with Different Data Amounts}
Previous methods need a large amount of clean data to mitigate backdoor, and thus, we want to explore the influence of the amount of clean data on backdoor mitigation. We compare mitigation results in Acc and ASR using Fine Pruning and ShapPruninig with different amounts of clean data in \cref{pruning on different dataset}. Our experiment illustrates that backdoor mitigation performance improves with the amount of clean data rising and there is a significant fluctuation in ASR during Fine Pruning with just 1 image per class. We attribute it to the lack of data to activate some normal neurons and there may be low activation values in many neurons, some of which are not poisoned. Furthermore, since data insufficiency weakens fine-tuning, Fine Pruning performance declines further. Similarly, ShapPruning with 300 images for each class performs best. But with the improvement of the data amount, backdoor mitigation is promoted to a relatively small extent. Furthermore our method performs the best in different experiments among all these defense methods with the same amount of data.

\subsection{Acceleration Comparison}
In this subsection, we compare our method of $\epsilon$-greedy with T-MAB \cite{ghorbani2020neuron} which uses Bernstein error bounds \cite{maurer2009empirical,mnih2008empirical} and find our method can more precisely and efficiently locate neurons with the largest top-$k$ Shapley values under the same average iterations. We estimate Shapley value of neurons with $50$ average iterations using these two methods. Meanwhile, we use the Monte-Carlo estimation of $5000$ average iterations as the actual Shapley value for this task. We arranged neurons randomly before $30$ average iterations and set $\epsilon$ to be $0.5$ and $0.3$ in $30$-$40$ and after $40$ iterations in $\epsilon$-greedy. We then compare these two methods' top-$50$ neurons and find them whether in the top-$70$ neurons in the MC experiment. Our experiments find that there are $46$ neurons in our methods' top-$50$ neurons in top-$70$ of actual value. On the contrary, there are only $27$ neurons found in T-MAB. We attribute the inaccuracy of T-MAB to that Bernstein error bound is too conservative and consumes too much time to determine which neurons' Shapley values are too small to calculate. On the contrary, our method combines exploration and utilization, getting more accurate estimations more efficiently.

\subsection{Data-free Backdoor Mitigation}
\label{data-free}
\begin{figure}
  \centering
  \begin{subfigure}{0.9\linewidth}
  \centering
  \includegraphics[width=\linewidth]{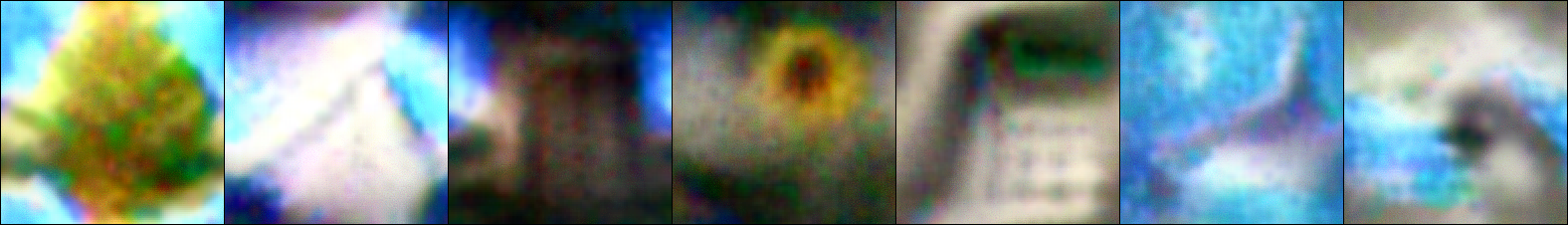}
  \end{subfigure}
  \hfill
  \begin{subfigure}{0.9\linewidth}
  \centering
  \includegraphics[width=\linewidth]{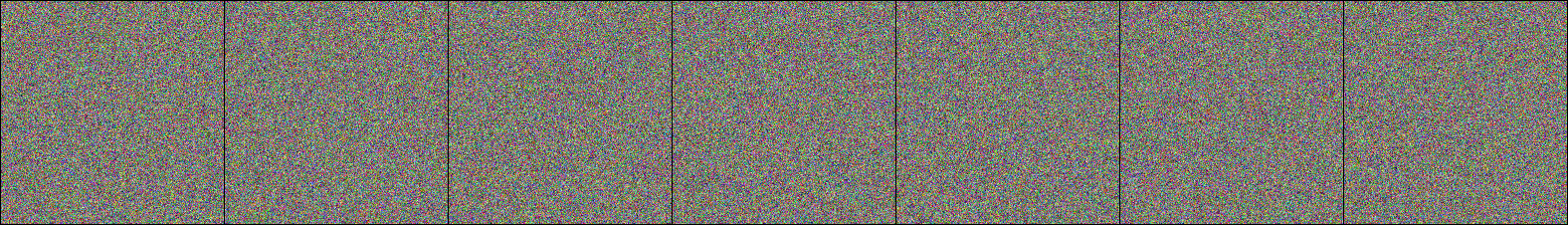}
  \end{subfigure}

\caption{Images recovered for ShapPruning (Ours) and DeepInspect. The first line is ours and the second line is DeepInspect.}
\label{Image-inversion}
\end{figure}

The experiment results with the few-shot setting mentioned above demonstrates that ShapPruning is robust against different attacks and architectures. Then, we further introduce our ShapPruning framework into a data-free situation. Firstly, we try to reverse the training images from the poisoned model and show them in \cref{Image-inversion}, where we compare our reversed images with model inversion attack \cite{fredrikson2015model} used in DeepInspect. Because DeepInspect's model inversion method is usually used in shallower networks \eg multilayer perceptron, the recovery results degrade sharply in VGG or ResNet. Furthermore, the similarity between the recovered images and real images influences trigger reverse and neuron activation, thus deciding backdoor defense performance. With the help of information in batch normalization layers, our method reconstructs better images.

Then we utilized the backdoor mitigation methods on the recovered images. Specifically, Fine Pruning (FP), Neural Cleanse (NC), and our ShapPruning method are conducted on our recovered images and DeepInspect (DI) is conducted with its original method. To test our methods' robustness, we evaluate on different attacks and architectures. We defend against BadNets on CIFAR10 and CIFAR100 with VGG16 and ResNet34 separately. Furthermore, we also defend on the different attack \eg Trojan Attack. The results are shown in \cref{tab:data-free}. With better-recovered images and robustness of mixture-mode ShapPruning, our method mitigates backdoor clearly with a small accuracy decline.

\section{Conclusions}
In this work, we propose Shapley Pruning framework to detect and mitigate backdoor attacks from poisoned models. Our method considers the interaction between neurons, locates the few infected neurons precisely, and protects models' structure and accuracy while pruning as many infected neurons as possible. Compared to prior work, our method mitigates backdoor successfully, using much fewer images (or even no clean data) and pruning much fewer neurons (about $1\%$ of total neurons) than previous methods. Furthermore, we mitigate backdoor with only less than $1\%$ accuracy decline in most situations. Also, our acceleration method, discarding threshold and $\epsilon$-greedy, can effectively reduce time consumption and help complete most tasks in just several minutes. Our method needs to reverse backdoor triggers for computing ASR in estimating Shaley value, which may be time-consuming. A more efficient and direct way is to use clean data for computing Shapley value to find the neurons in the model which contribute most to backdoor attacks, and we leave it to future work.
\section*{Acknowledgement} 
This work is partially funded by National Natural Science Foundation of China (Grant No. U21B2045, U20A20223) and Youth Innovation Promotion Association CAS (Grant No. Y201929).  Mr Zhuozhuo Tu is partially supported by ARC FL-170100117.

%%%%%%%%% REFERENCES
\newpage
{\small
\bibliographystyle{ieee_fullname}
\bibliography{egbib}
}

\end{document}